\documentclass[journal=nalefd,manuscript=letter, layout=standard]{achemso}

\usepackage{chemformula} % Formula subscripts using \ch{}
\usepackage[T1]{fontenc} % Use modern font encodings
\usepackage{graphicx}% Include Figure files
\usepackage{dcolumn}% Align table columns on decimal point
\usepackage{bm}% bold math
\usepackage{xcolor}
\usepackage{textgreek}
\newcommand*\dm[1]{#1$_2$}
\newcommand*\bias[1]{$U_\mathrm{DC} = #1$}
\newcommand*\current[1]{$I_\rm{t} = #1$} 
\renewcommand*\rm{\mathrm}

\author{Stepan Kovarik}
\affiliation{Department of Materials, ETH Zurich, Hönggerbergring 64, CH-8093 Zürich, Switzerland}
\email{stepan.kovarik@mat.ethz.ch}

\author{Roberto Robles}
\affiliation{Centro de Física de Materiales CFM/MPC (CSIC-UPV/EHU), Paseo Manuel de Lardizabal 5, 20018 San Sebastián, Spain}

\author{Richard Schlitz}

\author{Tom Sebastian Seifert}
\affiliation{Department of Materials, ETH Zurich, Hönggerbergring 64, CH-8093 Zürich, Switzerland}
\altaffiliation{Department of Physics, Freie Universität Berlin, 14195 Berlin, Germany}

\author{Nicolas Lorente}
\affiliation{Centro de Física de Materiales CFM/MPC (CSIC-UPV/EHU), Paseo Manuel de Lardizabal 5, 20018 San Sebastián, Spain}
\altaffiliation{Donostia International Physics Center (DIPC), Paseo Manuel de Lardizabal 4, 20018 San Sebastián, Spain}

\author{Pietro Gambardella}

\author{Sebastian Stepanow}
\affiliation{Department of Materials, ETH Zurich, Hönggerbergring 64, CH-8093 Zürich, Switzerland}
\email{sebastian.stepanow@mat.ethz.ch}
\title{Electron paramagnetic resonance of alkali metal atoms and dimers on ultrathin MgO}


\begin{document}

%\begin{tocentry}
%\end{tocentry}

\begin{figure}[ht]
\includegraphics{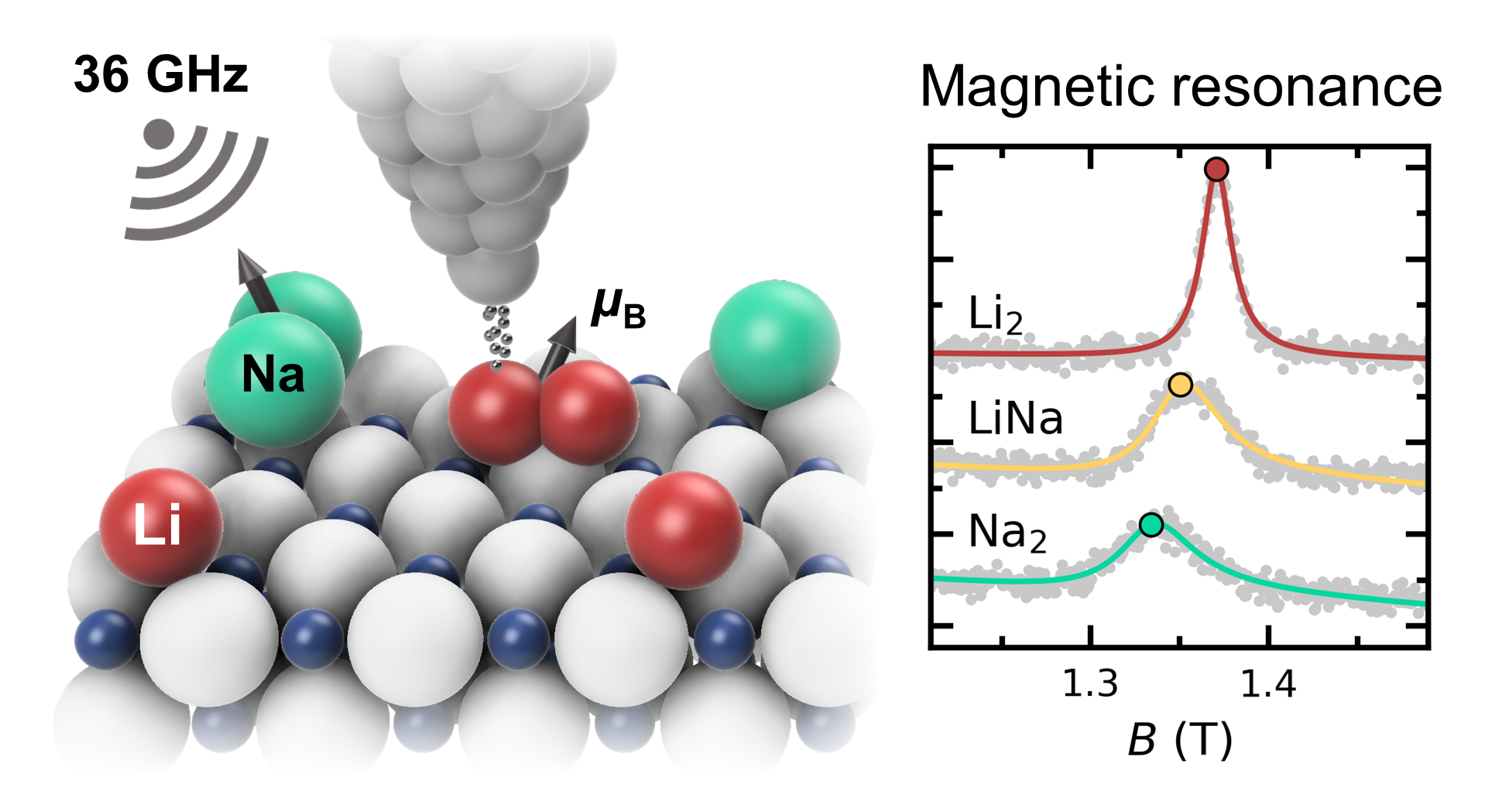}
\label{Fig:TOCGraphic}
\end{figure}
\newpage

\begin{abstract}
Electron paramagnetic resonance (EPR) can provide unique insight into the chemical structure and magnetic properties of dopants in oxide and semiconducting materials that are of interest for applications in electronics, catalysis, and quantum sensing. Here, we demonstrate that EPR in combination with scanning tunneling microscopy (STM) allows for probing the bonding and charge state of alkali metal atoms on an ultrathin magnesium oxide layer on Ag substrate. We observe a magnetic moment of $1\mu_\mathrm{B}$ for \dm{Li}, LiNa, and \dm{Na} dimers corresponding to spin radicals with a charge state of $+1e$. Single alkali atoms have the same charge state and no magnetic moment. The ionization of the adsorbates is attributed to charge transfer through the oxide to the metal substrate. Our work highlights the potential of EPR-STM to provide insight into dopant atoms that are relevant for the control of electrical properties of surfaces and nanodevices.
\end{abstract}

\section{Main text}
Recent studies have shown that electron paramagnetic resonance (EPR) spectroscopy can be combined with scanning tunneling microscopy (STM) to achieve single spin sensitivity and sub-nanometer spatial resolution \cite{Baumann2015, Choi2017, Seifert2020a, Seifert2020, Veldman2021, Steinbrecher2021}. This unique combination of techniques enables the investigation of the low-energy magnetic excitations of individual atoms on surfaces by overcoming the thermal resolution limit of tunneling spectroscopy. EPR-STM has been used to gain insight into the  interaction of the nuclear and electronic magnetic moments of single atoms \cite{Willke2018, Yang2018}, the dipolar \cite{Choi2017,Singha2021} and exchange coupling between atoms and molecules \cite{Yang2017, Bae2018, Zhang2022}, and the coherent spin dynamics of atoms and molecules on a surface \cite{Willke2018d, Yang2019b, Veldman2021, Willke2021}. These recent achievements highlight the potential of EPR-STM for the investigation of quantum phenomena on the atomic scale. However, the applicability of this technique to a broad range of systems remains an open question. Apart from early attempts to perform EPR spectroscopy on defect centers in Si \cite{Manassen1989, Balatsky2012}, which proved hard to reproduce, all of the EPR-STM studies mentioned above were performed on $3d$ transition-metal atoms, i.e., Fe, Ti, and Cu\cite{Baumann2015, Choi2017, Yang2017, Bae2018, Willke2018, Willke2018d, Yang2018, Yang2019b, Seifert2020a, Seifert2020, Steinbrecher2021, Veldman2021, Zhang2022, Singha2021} or an Fe ion in the iron phthalocyanine molecule\cite{Willke2021, Zhang2022} all adsorbed on the MgO surface. The application of this technique to other elemental systems and to a broader range of problems beyond single-atom magnetism remains to be demonstrated.

In this work, we show that EPR-STM can be applied to study the electronic configuration of alkali electron donor complexes on an oxide surface. Functionalized oxide surfaces are of particular importance for heterogeneous catalysis \cite{Freund2008}. One way of functionalization is to bring metal atoms to the surface of the oxides, which provide local binding sites and change in electron density and thus alter the chemical reactivity \cite{Pacchioni2013,Finazzi2008a}. A widely used system for fundamental studies in heterogeneous catalysis is MgO(100) with a small amount of metal atoms or clusters added to the bulk or surface \cite{Chiesa2007,Giordano2011}. The doping of MgO with alkali metals transforms the bare MgO substrate into a catalyst suitable for industry-relevant reactions, e.g., oxidative coupling of methane to produce higher order hydrocarbons \cite{Ito1985, Driscoll1985, Qian2020}. In such systems, the chemical transformation is thought to occur through the formation of electron-rich oxygen sites in the vicinity of the alkali atoms \cite{Driscoll1985}. However, the electronic state and chemical bonds of the dopant atoms, which are crucial for the catalytic activity, can only be determined by means of density functional theory (DFT) \cite{Finazzi2008a}.  
Here we use EPR-STM to investigate the charge state and binding configuration of individual Li and Na atoms as well as their dimers (\dm{Li}, \dm{Na}, and LiNa) deposited on ultrathin MgO(001) layers grown on a Ag substrate. Our measurements reveal a distinct EPR signal corresponding to a magnetic moment of $1$~\textmu$_\rm{B}$ for paramagnetic alkali metal dimers adsorbed near the top oxygen sites of MgO. In contrast, single alkali atoms adsorb on bridge O sites and show no magnetic moment. The absence of an EPR signal in the alkali atoms indicates a change of the orbital occupancy relative to the gas phase, where the valence $s$-shell is singly occupied. This change in occupancy shows that the atoms and dimers have the same charge state $+1e$ and that both transfer exactly one electron to the Ag substrate.
%, revealing that charge transfer is the relevant binding mechanism for alkali adatoms on ultrathin films of MgO. 
Our findings are supported by DFT calculations, which identify a single unpaired electron in the bonding molecular orbital as the origin of the magnetic moment of the alkali metal dimers. 

\begin{figure}[ht]
\includegraphics[keepaspectratio, width=8.45cm]{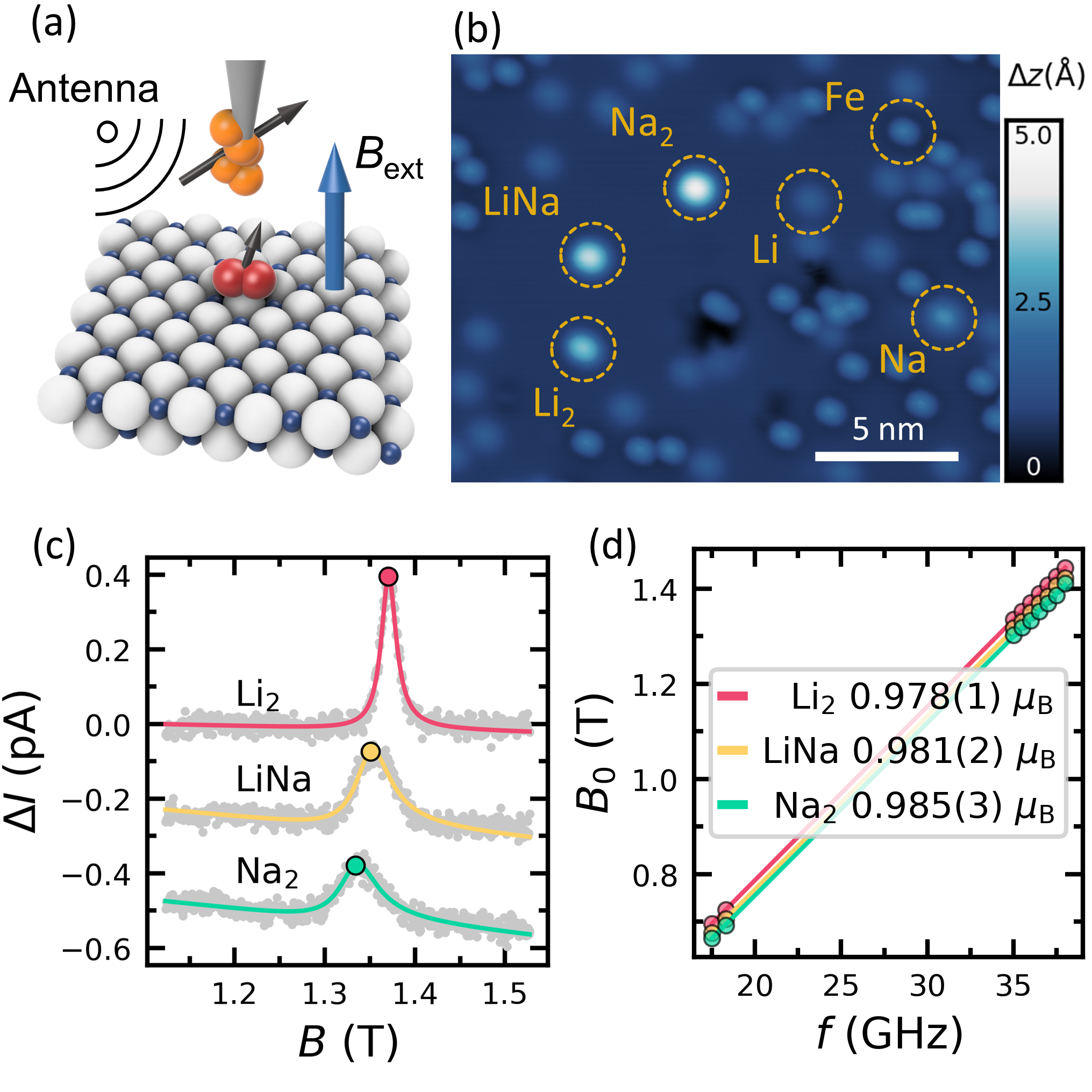}
\caption{\label{fig:F1} Alkali metal atoms and dimers on MgO. (a) Schematics of an alkali metal dimer (red) adsorbed on MgO under the magnetic STM tip decorated with Fe atoms (orange). Black arrows indicate the magnetic moment of the tip and the dimer. A voltage oscillating at radio-frequency $f$ is coupled to the tunnel junction to excite the EPR in an external magnetic field $B_\rm{ext}$. (b) Constant current STM image of alkali metal atoms and dimers on two-monolayers thick MgO on Ag(001). Examples of each species are labeled in yellow. Tunneling parameters: $U_\rm{DC} = 30$~mV, $I_\rm{t} = 50$~pA. (c) Typical EPR spectra observed on alkali metal dimers. The data points (gray) are fitted using a Fano lineshape (solid lines). The colored dots indicate the fitted resonance field $B_0$. The spectra are vertically offset for clarity. Tunneling parameters: $f = 36\, \rm{GHz}, U_\rm{rf} = 101\,\rm{mV},  U_\rm{DC} =-40\,\rm{mV}, I_\rm{t} = 40\,\rm{pA}$. (d) Dependence of $B_0$ on $f$ showing a linear relation typical for EPR. The magnetic moments of the dimers extracted from the slope are indicated in the legend. The error bars are smaller than the symbol sizes.}
\end{figure}

The EPR-STM experiments were performed in ultra-high vacuum with an STM operating at 4.5~K. In the STM, we applied a bias voltage $U_\rm{DC}$ to the sample and detected the tunneling current $I_\rm{t}$ by a transimpedance amplifier connected to the tip. The STM was equipped with a broadband (1-40~GHz) antenna that transmitted the signal with frequency $f$ from a microwave generator to the tunnel junction~\cite{Seifert2020} to resonantly excite the probed magnetic moment in the external magnetic field applied perpendicular to the sample surface [see Figure~\ref{fig:F1} (a)]. Two monolayers of MgO were grown by depositing Mg in an oxygen atmosphere ($10^{-6}\,$ mbar) on a clean Ag(001) surface heated to 690~K. 
%The Ag substrate was cleaned by cycles of sputtering with Ar$^+$ ions and annealing at 800~K. 
After the growth of MgO, the sample was inserted into the STM and cooled to 4.5~K, where Li, Na and Fe atoms were evaporated in sub-monolayer amounts. To achieve EPR sensitivity, we used spin-polarized tips prepared by transferring individual Fe atoms from MgO to the tip apex and applying an external magnetic field to polarize the magnetic cluster on the tip apex \cite{Baumann2015, Seifert2020a}. 
%Spin-polarization of the tip was confirmed with the observation of asymmetric inelastic excitation steps in the conductance spectrum of the Fe atoms.

The alkali metals adsorb mostly as single atoms on MgO [see Figure~\ref{fig:F1} (b)]. Individual atoms are imaged as round protrusions with an apparent height of 97(6)~pm and 197(10)~pm for Li and Na, respectively. 
%Note that we also deposited Cs atoms, which, however, moved on the surface while scanning due to the low adsorption energy \cite{Kim2014}. 
Less than 1~\% of the alkali atoms are found as dimers after deposition. The alkali metal dimers studied in this work were prepared by lateral manipulation of individual atoms with the STM tip stabilized at $U_\rm{DC} = 5$~mV and $I_\rm{t} = 5-8$~nA. Both the as-grown and the assembled dimers can be disassembled into individual atoms using the same settings. All dimers appear as nearly round protrusions with an apparent height dependent on their composition, i.e., \dm{Li} 260(6)~pm, LiNa 341(7)~pm, and \dm{Na} 390(9)~pm at $U_\rm{DC} = 30$~mV, $I_\rm{t} = 50$~pA. Those height variations enable reliable identification of each dimer.

To gain insight into the charge state and associated magnetic moment of the alkali adsorbates, we use EPR-STM in magnetic field sweep mode. To this end, we place a magnetic tip above a species of interest, apply a bias voltage oscillating at gigahertz frequency $f$  with peak amplitude $U_\rm{RF}$, and measure the time-averaged change in the tunneling current $\Delta I$ as a function of an external magnetic field $B_\rm{ext}$ \cite{Seifert2020}. Typical EPR spectra are shown in Figure~\ref{fig:F1}(c) for three different alkali metal dimers \dm{Li}, LiNa and \dm{Na}. EPR spectra recorded on individual alkali atoms (not shown) are featureless, suggesting the absence of a magnetic moment in the probed range from $0.87\,\mu_\mathrm{B}$ to $4.3\,\mu_\mathrm{B}$. The observed resonances of alkali dimers are fit by a Fano lineshape to extract the external resonant magnetic field $B_0$. 
Figure~\ref{fig:F1}(d) presents the dependence of $B_0$ on $f$, which allows us to extract the magnetic moment of the three alkali metal dimers. All configurations show a magnetic moment very close to 1~\textmu$_\rm{B}$ compatible with a spin $S=1/2$, i.e., the presence of one unpaired electron in the outer electronic shell with no orbital magnetic moment. We probe the magnetic moment only in the out-of-plane direction, however, we do not expect the strong anisotropy of magnetic moment due to the low symmetry of molecular orbital and weak spin-orbit interaction in alkali metals. The absence of an EPR signal for the single atom species suggests that the alkali atoms are stable on the MgO/Ag substrate in the ionized state $+1e$ with no electrons in the valence shell. 

The width of the EPR signals varies significantly between the different dimers. For the resonances presented in \ref{fig:F1}(c) the widths are 20.2(4)~mT, 47(2)~mT, and 51(2)~mT for \dm{Li}, LiNa, and \dm{Na}, respectively. Upon closer inspection of the spectral fits, we identify a minor but systematic deviation of the data from the Fano lineshape commonly used to fit EPR-STM data. The relative deviation of the fits is presented in\cite{SM}. All alkali metals have a non-zero nuclear magnetic moment $I=3/2$ for the most abundant $^7$Li (92.4\%) and $^{23}$Na (100\%) species \cite{Meija2016}, leading to a splitting of the EPR line of alkali atoms in gas phase into four lines with similar intensity. The separation of the lines is given by the coupling strength between the nuclear and electronic magnetic moment which is stronger for Na atoms  (885 MHz, 32 mT\cite{Chiesa2007}) compared to Li  (401 MHz, 14 mT \cite{Lian2008}) in the gas phase. Those coupling strengths are reduced upon adsorption on the MgO substrate by roughly 50\% \cite{Finazzi2008a}. Furthermore, we probe dimers instead of single atoms, which further decreases the coupling strength between the electronic and nuclear spins due to a reduced electron density of the bonding orbital at the position of the atomic nuclei \cite{Bruna2002}. Also, the interaction of the electron spin with two identical nuclei in the homodimers causes a splitting of the EPR line into seven equidistant lines with relative intensities following the ratio 1:2:3:4:3:2:1. In our setup the minimum line width that can be resolved is  $\approx$2 mT and we do not observe individual hyperfine-split resonant lines for the alkali dimers. However, the ratio of the atomic coupling strengths scales roughly as the observed line widths of the dimers. For the individual EPR transitions, we can only estimate the widths to be  between one and tens of mT. 

To further support this estimate, we have calculated the values of the isotropic hyperfine coupling constants following the method described in ref.\cite{Szasz2013} to be $a_{\text{Li,iso}} = 118$~MHz (4.2~mT) and $a_{\text{Na, iso}} = 209$~MHz (10.4~mT) for the alkali atoms in the homodimers on MgO using DFT. Using the calculated coupling strengths and assuming seven equally spaced resonant lines, we estimate the total splitting of the EPR resonance line to be $\approx 25.2$~mT in the case of Li and $\approx 62.4$~mT for Na. These numbers are in good agreement with the observed linewidths, taking into account the relative intensities of the EPR lines in homodimers.

\begin{figure}[ht]
\includegraphics[keepaspectratio, width=8.45cm]{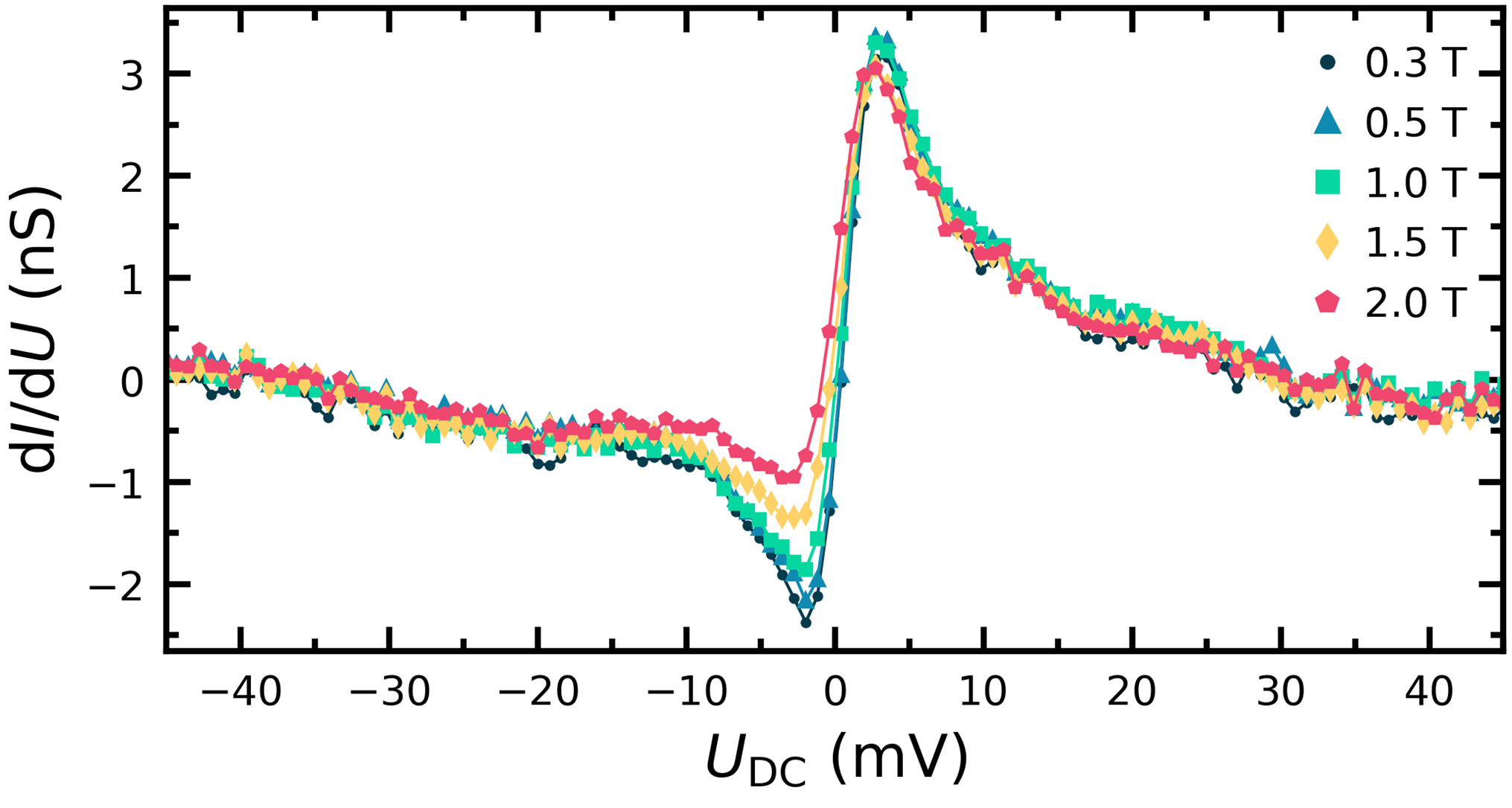}
\caption{\label{fig:F3} Differential conductance spectra of \dm{Li} as a function of external magnetic field. The spectra show a pronounced structure near zero bias arising from the spin-flip excitation of the magnetic moment with tunneling electrons. The spectra are measured with a spin-polarized tip stabilized above the center of \dm{Li} at $U_\rm{DC}=100$~mV and $I_\rm{t}=2$~nA. A spectrum obtained on bare MgO is subtracted from all spectra to remove features related to the electronic structure of the tip.} 
\end{figure}

The presence of a magnetic moment in the alkali metal dimers is confirmed by measurements of the differential conductance $\rm{d}I_\rm{t}/\rm{d}U_\rm{DC}$ of \dm{Li} as a function of external magnetic field, which also allows getting insight into the spin relaxation time through a quantitative analysis using a rate equation model presented in \cite{SM}. The $\rm{d}I_\rm{t}/\rm{d}U_\rm{DC}$ spectra presented in Figure~\ref{fig:F3} show two noticeable features: a prominent negative-positive peak around zero bias and a magnetic field dependence of the peak at negative bias. We attribute the zero-bias structure to the spin-flip excitations of the magnetic moment of the dimer caused by the spin-polarized tunneling electrons, in analogy with previous measurements of transition-metal atoms on insulating substrates \cite{Heinrich2004, Gauyacq2012, Yang2017}. 
The spin-polarization of the tip causes different excitation probabilities depending on the tunneling current direction, that leads to different conductance step heights for negative and positive bias \cite{Loth2010e, Loth2010b}. The change in the polarisation of the tip with the applied magnetic field induces the change in the intensity of the dip at the negative bias in the data presented in \ref{fig:F3}. The dip and peak shape at negative and positive bias, respectively, are due to additional saturation effects of the spin-flip excitations that are observed in systems where the spin relaxation time is comparable to the average time between tunneling events \cite{Ternes2015, Loth2010b}. Based on the quantitative analysis of the evolution of the conductance spectra presented in \cite{SM}, we estimate the energy relaxation time $T_1\approx 2$~ns for the experimental parameters used to perform the EPR-STM measurements presented in Figure~\ref{fig:F1}(c).

\begin{figure}[ht]
\includegraphics[keepaspectratio, width=8.45cm]{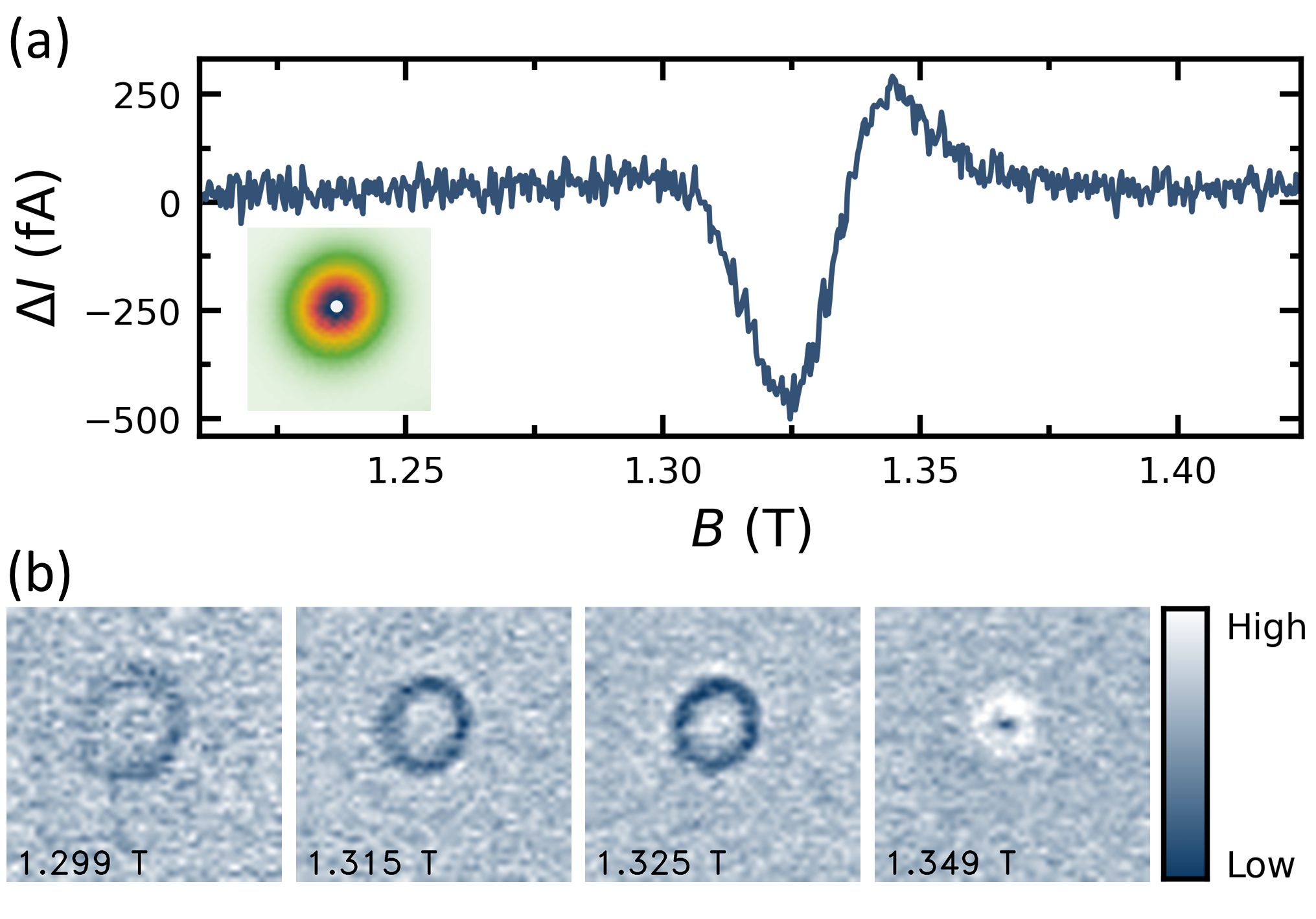}
\caption{Spatial distribution of the EPR signal of \dm{Li}. (a) EPR spectrum measured with frequency modulation. The inset shows a constant height image of \dm{Li}. The white dot marks the tip position during measurement of the EPR spectrum. (b) Spatial map of the EPR signal at four different external magnetic fields. The tip was stabilized above the center of the dimer at \current{40}~pA and \bias{-50}~mV before acquisition of each image (b) and spectrum (a). The data are measured with a modulation amplitude of 128 MHz at 36.051 GHz. All scans are $1.5\times1.5$ nm$^2$. \label{fig:F2}}
\end{figure}

Besides probing unpaired electrons, EPR-STM also provides information on the spatial distribution of the spin density around the adsorbates. Figure~\ref{fig:F2} shows a spectrum and the spatial distribution of the EPR signal of \dm{Li} measured using frequency modulation of the oscillating voltage. In contrast to the amplitude modulation used in Figure~\ref{fig:F1}, frequency modulation allows us to obtain a spatial map of the EPR signal without the need to subtract a reference image to suppress the signal caused by the rectification of the excitation voltage \cite{Seifert2020}. Moreover, the bipolar shape of the EPR signal recorded with the frequency modulation allows us to distinguish tip positions in which the probed species senses a magnetic field that is higher or lower than the resonant field. Since the local magnetic field at the position of the probed species is given by the external magnetic field and by interaction with the magnetic tip, the variation of the tip position causes detuning from the resonance \cite{Wilke2019}. This is shown in the spatial maps of the EPR signal in Figure~\ref{fig:F2}(b) obtained at several magnetic fields close to the resonant field. The EPR signal is distributed in rings with high values on the outer edge and low values toward the center of the ring. This behavior is due to a decreased energy splitting of the spin states as the tip approaches the center of the dimer, i.e. a shift of the EPR spectrum to higher external magnetic field values. Thus, the magnetic tip couples antiferromagnetically to the surface spin. From the almost circular shape of the signal, we conclude that the interaction of the tip with the dimer is dominated by the isotropic exchange interaction between the tip states and a nearly spherical orbital. This observation is consistent with the unpaired electron of \dm{Li} residing in a bonding orbital.

\begin{figure}[ht]
\includegraphics[keepaspectratio, width=8.45cm]{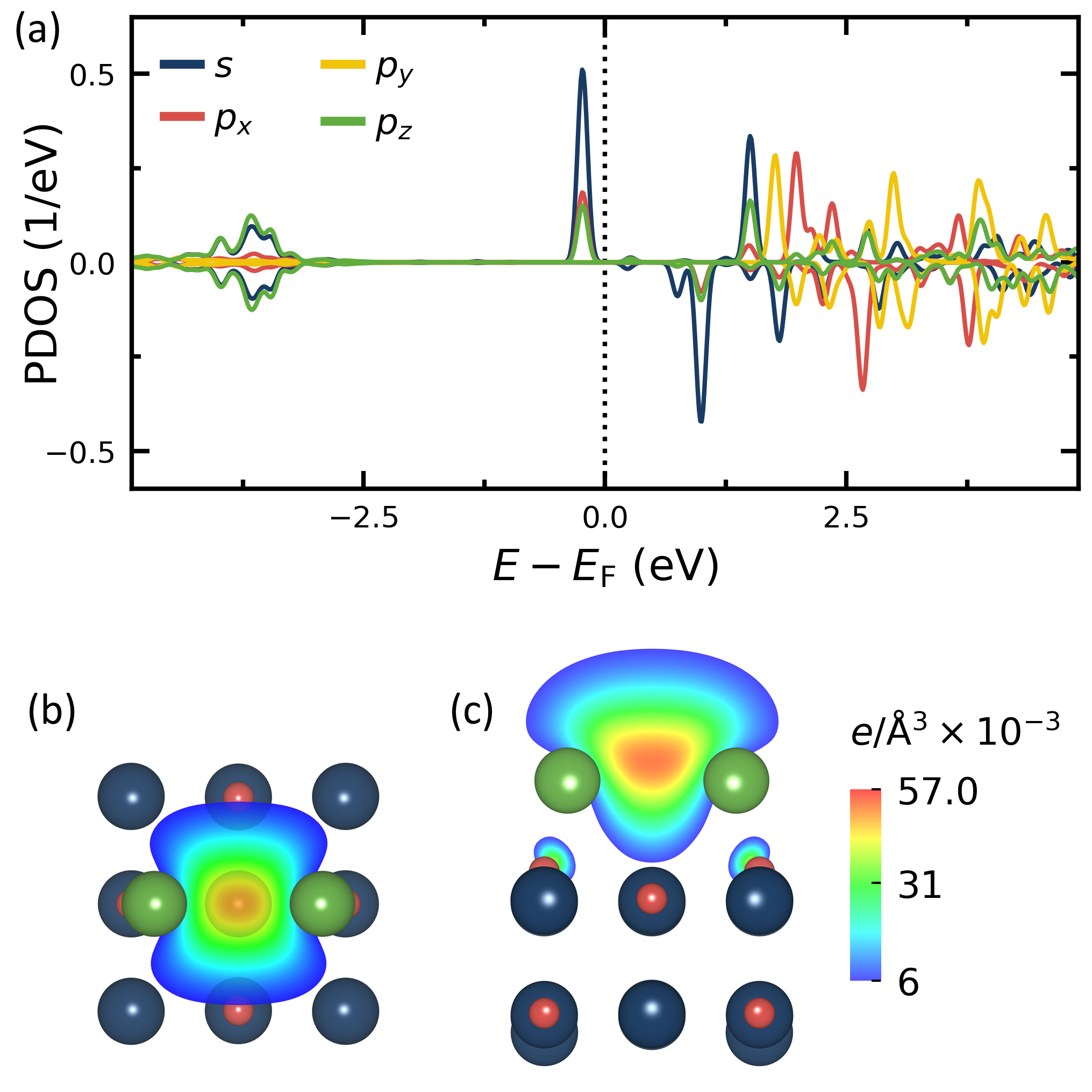}
\caption{\label{fig:F4} (a) Projected DOS \dm{Li} on MgO calculated by DFT. Projections of the local DOS on the different valence $s-$ and $p$-orbitals are labeled by colors. (b) Top view and (c) side view along (010) of the binding configuration of \dm{Li}. Color maps show the spin density distribution on orthogonal planes intersecting \dm{Li}. Color code of the atoms in (b) and (c) is Mg: blue, O: red, and Li: green} 
\end{figure}

To gain further insight into the binding and electronic configurations of the alkali adsorbates, we performed DFT calculations of Li and Na atoms and dimers on MgO/Ag(001). We find that the most stable binding site of Li and Na monomers is the bridge site of the oxygen sub-lattice. In contrast, the alkali atoms in the dimers bind near the oxygen top sites, with the axis of the dimer oriented diagonally across the unit-cell of the oxygen sub-lattice. The calculated binding configurations agree with STM measurements \cite{SM}. The calculated bond length of \dm{Li}/MgO is stretched by 3\% compared with the ionic molecules in the gas phase \cite{DeSousa2019}. In contrast, the bond lengths of LiNa and \dm{Na} are compressed by 11\% and 6\%, respectively. The adsorption energies of the dimers are: 2.24~eV, 1.68~eV, and 1.08~eV for \dm{Li}, LiNa, and \dm{Na}, respectively. 

The DFT calculations of individual Li and Na atoms identify a charge of around $+1e$ and no magnetic moment. The same charge state has been predicted earlier for K atoms adsorbed on 2 ML of MgO/Ag(001) \cite{Giordano2006}. On ultrathin MgO the alkali atoms are positively charged due to the transfer of one electron to the nearby metal substrate via tunneling through the oxide \cite{Freund2007, Giordano2011}. This behavior is remarkably different from that of alkali atoms on the surface of bulk MgO, which maintain a neutral charge state and adsorb directly above oxygen \cite{Lian2008}. 

In general, adsorbates on an oxide surface have a modified electronic structure that is affected by hybridization with the orbitals of the substrate, charge transfer, electrical polarization effects including dispersive forces as well as lattice deformations. For alkali metals adsorbed on bulk MgO, polarization effects dominate. Therefore, the alkali atoms remain electronically neutral but the electrical polarization modifies the orbital structure of the atom. The original spherically symmetric $s$ orbital becomes polarized by mixing with $p_z$ orbitals \cite{Pacchioni2013,Chiesa2007}. Our EPR maps and DFT calculations show that this does not happen for the ionized alkali dimers on ultrathin MgO, indicating that both the charge state and the binding site of the alkali species depend on the thickness of the oxide layer.  

Unlike individual atoms, the calculated spin-dependent density of states (DOS) of the alkali metal dimers reveal a significant spin polarization of the electronic states close to the Fermi level. Figure~\ref{fig:F4}(a) shows the calculated DOS of Li in \dm{Li}. The peak in the DOS just below the Fermi level corresponds to the bonding molecular orbital that arises from the hybridization of the $2s-$ and $2p-$orbitals of the two Li atoms. This orbital is occupied by one electron, which gives rise to a finite spin-density [see Figure~\ref{fig:F4}(b-c)]. In the gas phase, the \dm{Li} molecule is electrically neutral and has no magnetic moment. The DFT results thus agree with the experimental observation of a magnetic moment of 1~\textmu$_\rm B$ and are consistent with the electropositive character of the alkali metals, which leads to a charge transfer from \dm{Li} to the substrate resulting in the charged state $+1e$. 

Both homodimers \dm{Li} and \dm{Na} have the same magnetic moment as the heterodimer LiNa, indicating the same charge state. In contrast, the spin distribution of the heterodimer is qualitatively different compared to the homodimers, since it lacks mirror symmetry and has a maximum spin density closer to the Li atom \cite{SM}. This asymmetry has not been resolved in the experiment, as we do not observe any prominent anisotropic feature of the electron density around the dimers in the STM images. This is attributed to the small spatial extension and almost round profile of the bonding orbital above the dimer [see Figure~\ref{fig:F4}(c)].

Finally, we note that the synthesis of alkali trimers and more complex arrangements could not be achieved, probably due to a low dissociation energy of larger clusters. Preparing larger assemblies causes the immediate splitting of trimers into a dimer and a single atom, which further underscores the higher stability of the singly-charged dimers.

Our work sheds light on the bonding and charge state of alkali metal adsorbates on ultrathin MgO/Ag(001) and opens a new range of systems for EPR-STM investigations beyond $3d$ transition metals \cite{Baumann2015, Choi2017, Yang2017, Bae2018, Willke2018, Willke2018d, Yang2018, Yang2019b, Seifert2020a, Seifert2020, Steinbrecher2021, Veldman2021, Zhang2022, Singha2021, Seifert2021}. The ability to probe electron donor species using EPR-STM may be useful to perform atomic-scale characterization of silicon-based spin qubits \cite{Mueller2021, Morello2020} and deterministic manipulation of dopants in semiconductors \cite{Koenraad2011, Fuechsle2012, Mueller2021} and molecular adsorbates \cite{Krull2013}. Other systems of interests are alkali metal doped graphene \cite{Ludbrook2015} and organic semiconductors \cite{Tanigaki1991, Mitsuhashi2010}, where charge transfer from the different alkali metal species induces a superconducting phase transition. 

\begin{acknowledgement}
We acknowledge funding from the Swiss National Science Foundation, Project No. 200021\_163225. RR and NL thank Grant RTI2018-097895-B-C44 funded by MCIN/AEI/10.13039/501100011033 and by ''ERDF A way of making Europe''.
\end{acknowledgement}

\begin{suppinfo}
Additional EPR-STM data, measurement of EPR-STM on \dm{Li} at different experimental parameters, analysis of differential conductance spectra of \dm{Li}, identification of the binding configurations of alkali metals, additional details of DFT calculations.
\end{suppinfo}

\bibliography{lit}
\end{document}